\documentclass[a4paper,12pt]{article}
\begin{document}
\sf
{\noindent \large Partial-wave Coulomb $t$-matrices for like-charged\\
particles at ground-state energy } \\[.2in]
{\sf V. F. Kharchenko}\\[.1in] 
{\footnotesize Bogolyubov Institute for Theoretical Physics,  
National Academy of Sciences \\of Ukraine, UA - 03143, Kyiv, Ukraine \\ [.1in]
E-mail: vkharchenko@bitp.kiev.ua} \\[.1in]
\noindent \small{ {\sf Abstract }  \\ 
{\small We study a special case at which the analytical solution of the 
Lippmann-Schwinger integral equation for the partial wave two-body Coulomb 
transition matrix for likely charged particles at negative energy is possible. 
With the use of the Fock's method of the stereographic projection of the 
momentum space onto the four-dimensional unit sphere, the analytical expressions 
for s-, p- and d-wave partial Coulomb transition matrices for repulsively
interacting particles at bound-state energy have been derived.} \\ [.05in]

\noindent {\sf 1. Introduction} \\ 

\noindent The Coulomb transition matrix ($t$-matrix), being a scalar function of the 
initial and final relative momenta and the energy, provides all information about 
the system of two interacting charged particles. The analytic properties of the Coulomb 
$t$-matrix have been discussed in the review [1]. The availability of bound states for 
systems with oppositely charged particles leads to the appearance of pole singularities 
of the corresponding Coulomb $t$-matrix at bound-state energies with residues relating 
with wave functions of the bound states. In the case of likely charged particles the 
Coulomb $t$-matrix  has no energy poles. The analytic properties of the Coulomb $t$-matrix,  
which in the case of short-range interaction potentials manifest themselves as a singular 
branch point with a cut along the positive energy axis and related unitarity conditions 
on and off the energy shell, are more complicated (see review [1]).

The knowledge of the two-body Coulomb transition matrix with the momenta off 
the energy shell is especially important when studing properties of few=body atomic and 
nuclear systems containing charged particles with the use of the Faddeev [2,3] and 
Faddeev-Yakuboskii [4] integral equations. For such systems, Faddeev equations are known 
to become non-Fredholm even below the decay threshold. The extraction of the main Coulomb 
singularity and the regularization of three-body equations in this case were proposed by 
Veselova [5] with the help of the known Gorshkov procedure for two-body systems [6]. 
The problem of regularization of the integral equations for four-body systems containing 
charged particles was considered in work [7]. Earlier information relative to the 
properties of the two-body off-shell Coulomb transition matrix can be found in [1]. 

Several representations for two-body Coulomb transition matrix are known in the 
literature [8-16]. Of special interest is the study of the Coulomb transition matrix 
with the use of the Coulomb system symmetry in the Fock four-dimensional Euclidean 
space [17]. Earlier, the Fock method was applied in Bratsev-Trifonov's[10] and 
Schwinger's [12] works in order to derive the Coulomb Green's function in the 
one-parameter integral form. Expressions for the three-dimensional Coulomb transition 
matrix with explicitly singled out transfered momentum and energy singularities were 
obtained in works [15] (for negative energies, $E<0$) and [16] (for zero and positive 
energies,  $E\geq 0$).

For the first time, a possibility to derive an analytical expression fortial wave 
two-body Coulomb transition matrices at the ground bound state energy was examined 
for oppositely charged particles (with the attractive interaction) in the previous 
work [18]. In this work, on the basis of the Fock method of stereographic projection 
of the three-dimensional momentum space onto a four-dimensional unit sphere [17], 
the form of the partial wave Coulomb transition matrices for a system of two likely 
charged bodies (with the repulsive Coulomb interaction) is analyzed. The consideration 
begins in Section 2, where the expression obtained earlier in work [14] for the 
three-dimensional Coulomb transition matrix at the negative energy is used. 
In Section 3, a general expression for the off-shell partial wave Coulomb $t$-matrix 
at the negative energy is derived. Section 4 is devoted to the study of the partial 
wave Coulomb $t$-matrix at the ground bound state energy, and it is shown that a simple 
analytical expression for partial wave $t$-matrix can be obtained in this case. 
Explicit analytical expressions for the s-, p- and d-wave components of the 
Coulomb $t$-matrix are presented. Final remarks and conclusions are made 
in Section 5.\\ [.1in] 

\noindent {\bf 2. Three-dimensional Coulomb transition matrix at the negative \\ 
energy with explicitly separated singularities }\\ 

\noindent The three-dimensional Coulomb transition matrix $\langle {\bf k}|t(E)|
{\bf k}^{\prime}\rangle$ stisfies the inhomogeneous Lippmann-Schwinger integral 
equation 
\begin{equation}
<{\bf k}|t(E)|{\bf k}^{\prime}>=\langle {\bf k} \mid v \mid {\bf k}^{\prime}\rangle + 
\int \frac{d{\bf k}^{\prime\prime}}{(2\pi)^3}\langle {\bf k} \mid v \mid 
{\bf k}^{\prime\prime}\rangle \frac{1}{E-\frac{k^{\prime\prime 2}}{2\mu}} 
<{\bf k}^{\prime\prime}|t(E)|{\bf k}^{\prime}>\;\;.   
\end{equation}
Here, the free term $\langle {\bf k} \mid v \mid {\bf k}^{\prime}\rangle $ 
is determined by the Coulomb interaction potential $v(r)=q_1 q_2 / r$, where 
$q_i$ is the charge of the $i$-th particle ($i=1, 2$), and $r$ is the distance 
between particles 1 and 2. In the momentum space, this term looks like
\begin{equation}
\langle {\bf k} \mid v \mid {\bf k}^{\prime}\rangle = 
\frac{4\pi q_1 q_2}{\mid {\bf k}-{\bf k}^{\prime}\mid ^2}\;. 
\end{equation}
where ${\bf k}$ and ${\bf k}^{\prime}$ are relative momenta corresponding to the 
radius-vectors ${\bf r}$ and ${\bf r}^{\prime}$, respectively, in the coordinate 
space. The kernel of the integral equation (1) is a product of the operator of 
Coulomb interaction potential (2) and the free Green operator
\begin{equation}
<{\bf k}|g_0(E)|{\bf k}^{\prime}>=\frac{(2\pi)^3 \delta ({\bf k}-{\bf k}^{\prime})}
{E-\frac{k^2}{2\mu}}\;,
\end{equation}
where the quantity $E$ is the total energy of the relative motion of particles 
1 and 2, and $\mu = m_1 m_2/(m_1 + m_2)$ is their reduced mass. 

In this work, the consideration is confined to the problem of Coulomb scattering 
of two likely charged particles off the energy shell in case of negative energy
\begin{equation}
E = - \frac{\hbar^2 \kappa^2}{2\mu}\;.
\end{equation} 
The consideration is based on the solution on the integral equation (1) for the 
off-shell three-dimensional Coulomb transition matrix with the explicitly 
separated singularities in the variables of the transfer momentum and the energy, 
which was obtained by us early [14]: \pagebreak
\begin{displaymath}
 <{\bf k}|t(E)|{\bf k}^{\prime}>=\frac{8\pi q_1 q_2 \kappa^2}{(k^2+\kappa^2)
(k^{\prime 2}+\kappa^2)\sin \omega}\left[ \cot \frac{\omega}{2} 
-\pi\gamma \cos\gamma\omega 
-\gamma \sin 2\gamma \omega \ln \left( \sin\frac{\omega}{2} \right) \right.
\end{displaymath}
\begin{equation}
 + 2\pi\gamma\; c(\gamma)\; \cot \gamma\pi \sin\gamma\omega 
+  \gamma \cos \gamma \omega \int_{0}^{\omega} d\varphi \;
\sin \gamma\varphi\; cot\frac{\varphi}{2} 
\end{equation} 
\begin{displaymath}
\left.  + 2\gamma^2 \sin\gamma\omega 
\int_{\omega}^{\pi} d\varphi \; \sin \gamma\varphi \;\ln \left( (\sin \frac{\varphi}{2} 
\right)\right]\;,
\end{displaymath}
where $\gamma$  
\begin{equation}
\gamma = \frac{\mu q_1 q_2}{\hbar^2 \kappa}\;,
\end{equation} 
is the dimensionless Coulomb parameter, and $\hbar$ is the reduced Planck's 
constant. The variable $\omega$ in Eq.(5) stands for the angle between two 
4-dimensional unit vectors $e\equiv ({\bf e}, e_0)$ and $e^{\prime}\equiv 
({\bf e^{\prime}}, e_0^{\prime})$ in the four-dimensional Euclidean space  
introduced by Fock [17]: 
\begin{equation}
{\bf e}=\frac{2\kappa {\bf k}}{\kappa^2 + k^2}\;,\quad 
e_0 = \frac{\kappa^2 - k^2}{\kappa^2 + k^2} \quad  \mbox{  ³  } \quad
{\bf e}^{\prime}=\frac{2\kappa {\bf k}^{\prime}}{\kappa^2 + k^{\prime 2}}\;,\quad 
e_0^{\prime} = \frac{\kappa^2 - k^{\prime 2}}{\kappa^2 + k^{\prime 2}}\;; 
\end{equation} 
\begin{equation}
\cos \o^{\prime} = {\bf e}\cdot {\bf e}^{\prime} +  e_0\cdot 
e_0^{\prime}\;.
\end{equation} 
The three-dimensional vectors ${\bf k}$ and ${\bf k}^\prime$ lie in a hyperplane, 
which is a stereographic projection of a sphere with the unit radius. The variable   
$\omega$ is determined by the relation 
\begin{equation}
\sin^2\frac{\omega}{2} = \frac{{\kappa^2}\mid {\bf k} - {\bf k}^\prime \mid ^2}
{(k^2 + \kappa^2)(k^{\prime 2} + \kappa^2)}\;\; , \;\; 0\leq \omega \leq \pi\;\;.
\end{equation} 
The function $c(\gamma)$ in Eq.(5) looks like
\begin{equation}
c(\gamma)= \frac{1}{2} \left( 1 - \frac{1}{\pi}\int_{0}^{\pi} d\varphi \;
\sin \gamma\varphi 
\;\cot \frac{\varphi}{2} \right)
\end{equation}
or in terms of the digamma functions
\begin{equation}
c(\gamma)= \theta(-\gamma)+\frac{\sin \gamma\pi}{2\pi} \left[ \psi \left( 
\frac{\mid \gamma\mid +1}{2} \right)- \psi \left( \frac{\mid \gamma\mid }{2} 
\right) - \frac{1}{\mid \gamma\mid } \right]\;,
\end{equation}
where $\psi(x)\equiv d/dx \ln\Gamma(x)$ and $\Gamma(x)$ are the digamma- and 
gamma-functions [18], and $\theta(x)$ is the Heaviside step function,
\begin{displaymath}
\theta(x)=\left\{ \begin{array}{cc}
                 1 & \mbox{for $x>0$}\;, \\
                 0 & \mbox{for $x<0$}\;.
                 \end{array}
          \right.
\end{displaymath}

The first three terms in the square brackets in Eq.(5) contain transferred 
momentum singularities
\begin{displaymath}
\mid {\bf k}-{\bf k}^{\prime}\mid ^{-2}\;, \quad \mid {\bf k}-{\bf k}^{\prime}\mid ^{-1} \;
\mbox{ and }\; \ln \left\{\kappa \mid {\bf k}-{\bf k}^{\prime}\mid 
/(k^2 + \kappa^2)^{1/2}(k^{\prime 2} + \kappa^2)^{1/2}\right\}\;,
\end{displaymath}
respectively. The other three terms in Eq.(5) are smooth functions of 
$\mid {\bf k}-{\bf k}^{\prime}\mid$.

The fourth term in expression (5) contains singularities in the energy. 
They arise only in the case of attractive Coulomb potential (with opposite 
electric charges, $q_1 q_2 < 0$), when the Coulomb parameter $\gamma$ 
accepts negative integer values corresponding to the spectrum of bound 
states of a two-particle system with the energies $E=E_n$,
\begin{equation}
E_n = - \frac{\mu (q_1 q_2)^2}{2 \hbar^2 n^2}\;,\qquad n = 1, 2, 3,\cdots \;.
\end{equation}
According to EQS. (4) and (6), the corresponding values of the parameter 
$\kappa$ and the Coulomb parameter $\gamma$ are equal to 
\begin{equation}
\kappa_n = \frac{\sqrt{-2\mu E_n}}{\hbar} = \frac{\mu \mid q_1 q_2 \mid }
{\hbar^2 n}\;,\qquad \gamma_n = \frac{\mu q_1 q_2 }{\hbar^2 \kappa_n} = 
\frac{q_1 q_2}{\mid q_1 q_2 \mid} n\;,
\end{equation}
respectively. At these points $\gamma=\gamma_n = -n$, so that the function 
$\cot \gamma\pi$ has pole singularities, and the function $c(\gamma)$ 
differs from zero: $c(-n) = 1$. 

In this case of repulsive Coulomb potential ($\gamma$>0), the expression 
for $c(\gamma)$ equals zero at positive values of $\gamma$, $c(n) = 0$, 
and the fourth term in Eq. (5) is finite and equal to 
\begin{equation}
\left. \rho(\gamma) \equiv \frac{2\pi\gamma c(\gamma)}{\tan \gamma\pi}
\right|_{\gamma \rightarrow n} = \;\; \rho_n\;,
\end{equation}
where
\begin{equation}
\rho_n =2n c^{\prime}(n), \qquad c^{\prime}(n)=-\frac{1}{2\pi} \int_0^\pi d\varphi\; 
\cos n\varphi \;\cot \frac{\varphi}{2} 
\end{equation}
or, using the function $\beta(x) = \frac{1}{2}\left[ \psi \left( \frac{x+1}
{2} \right)- \psi \left( \frac{x}{2} \right) \right]$,
\begin{equation}
\rho_n = (-1)^n \left[ 2 n \beta (n) - 1 \right]\;.
\end{equation}
The ultimate expression for $\rho_n$ takes the form 
\begin{equation}
\rho_n = (-1)^n - 2 n \ln 2 - 2 n \sum_{m=1}^{n} \frac{(-1)^m}{m}\;.
\end{equation}
\\ [.1in]  

\noindent {\bf 3. Partial wave component of the Coulomb transition matrix \\
at negative energy }\\ 

\noindent Using the partial wave method and expanding the matrix elements 
of the Coulomb potential and the transition matrix with a negative energy 
in series in Legendre polinomials $P_l(x)$,
\begin{displaymath}
\langle {\bf k} \mid v \mid {\bf k}^{\prime}\rangle = \sum_{l=0}^{\infty} 
(2l+1) v_l(k,k^{\prime}) P_l(\hat{{\bf k}}\cdot \hat{{\bf k}}^{\prime})\;,
\end{displaymath}
\begin{equation}
\langle {\bf k} \mid t(E) \mid {\bf k}^{\prime}\rangle = \sum_{l=0}^{\infty} 
(2l+1) t_l(k,k^{\prime};E) P_l(\hat{{\bf k}}\cdot \hat{{\bf k}}^{\prime})\;,
\end{equation}
where $\hat{\bf k}$ is a unit vector along the vector ${\bf k}$, and 
$\hat{\bf k}\cdot \hat{{\bf k}}^{\prime}=\cos \theta$, the one-dimensional 
integral equation for the partial wave component of the transition matrix 
can be written in the form  
\begin{equation}
t_l(k,k^{\prime};E) = v_l(k,k^{\prime})+
\int_{0}^{\infty} \frac{dk^{\prime\prime} {k^{\prime\prime}}^2}{2\pi^2}
v_l(k,k^{\prime\prime}) \frac{1}{E-\frac{{k^{\prime\prime}}^2}{2\mu}}
t_l(k^{\prime\prime},k^{\prime};E)\;\;. \\[3mm]  
\end{equation}  
The inhomogeneous and the kernel of this equation contain a partial wave 
component of the Coulomb interaction potential 
\begin{equation}
v_l(k,k^{\prime}) = \frac{1}{2} \int_{0}^{\pi} d\theta \;\sin \theta \;
P_l(\cos \theta)
\langle {\bf k} \mid v \mid {\bf k}^{\prime}\rangle \;.
\end{equation}

According to definition (18), the partial wave component of the Coulomb
transition matrix $t_l(k,k^{\prime};E)$ equals 
\begin{equation}
t_l(k,k^{\prime};E) = \frac{1}{2} \int_{0}^{\pi} d\theta \;\sin \theta \;
P_l(\cos \theta)\;\langle {\bf k} \mid t (E) \mid {\bf k}^{\prime}\rangle \;.
\end{equation}

Taking into account that expression (5) for the three-dimensional 
Coulomb transition matrix $\langle {\bf k} \mid t (E) \mid 
{\bf k}^{\prime}\rangle$ depends on the angle $\omega$ between 
the unit vectors $e$ and $e^{\prime}$ in the four-dimensional Fock space, 
it is convenient to go in Eq. (21) from the integration over the angle
$\theta$ between the vectors $\hat{\bf k}$ and $\hat{\bf k}^{\prime}$ to 
the integration over the angle$\omega$. From expression (9) describing the 
relationship between the angles $\theta$ and $\omega$, it follows that
\begin{equation}
\cos {\theta} = \frac{\xi}{\eta} - \frac{1}{\eta} {\sin}^2{\frac{\omega}{2}} 
= \frac{2\xi - 1 + \cos {\omega}}{2\eta}\;,
\qquad \qquad \sin {\theta} \; d\theta = \frac{1}{2\eta} \sin {\omega} \; 
d\omega \;,   
\end{equation}    
where
\begin{equation}
\xi = \frac{\kappa^2 (k^2 + {k^{\prime}}^2)}{(k^2 + \kappa^2)({k^{\prime}}^2 
+ \kappa^2)}\;,\quad \eta = \frac{2 \kappa^2 k k^{\prime}}{(k^2 + 
\kappa^2)({k^{\prime}}^2 + \kappa^2)}\;.
\end{equation}  
Then the formula (21) can be rewritten as
\begin{equation}
t_l(k,k^{\prime};E) = \frac{1}{4\eta} \int_{\omega_0}^{\omega_{\pi}} 
d\omega\; \sin \omega \; P_l \left( \frac{2\xi-1+ 
\cos \omega}{2\eta} \right)
\langle {\bf k} \mid t (E) \mid {\bf k}^{\prime}\rangle  \;.
\end{equation}
The integration limits in Eq. (24) are determined by the expressions
\begin{equation}
\omega_0 = 2 \arcsin \sqrt{\xi - \eta}\;, \qquad \; \omega_{\pi} 
= 2 \arcsin \sqrt{\xi + \eta}\;,
\end{equation} 
so that
\begin{displaymath}
\cos \omega_0 = 1 - 2 \xi + 2 \eta\;, \qquad \; \cos \omega_{\pi} 
= 1 - 2 \xi - 2 \eta\;,
\end{displaymath}
\begin{equation}
\sin \omega_0 = 2 \sqrt{\xi - \eta} \sqrt{1 - \xi + \eta}\;, \qquad \; 
\sin \omega_{\pi} = 2 \sqrt{\xi + \eta} \sqrt{1 - \xi - \eta}\;.
\end{equation} 

Substituting expression (5) for the three-dimension transition matrix 
int  Eq. (24), we obtain the following formula for the partial wave 
Coulomb transition matrix $t_l(k,k^{\prime};E)$ at $E<0$: 
\begin{displaymath}
t_l(k,k^{\prime};E)= \frac{\pi q_1 q_2}{k k^{\prime}} 
\int_{\omega_0}^{\omega_{\pi}} d\omega \;P_l \left( \frac{2\xi-1+ 
\cos \omega}{2\eta}\right) \left\{ \cot {\frac{\omega}{2}} \right.
\end{displaymath}
\begin{equation}
- \pi \gamma \; \cos \gamma \omega - \gamma \;\sin 2\gamma\omega \;
\ln (\sin \frac{\omega}{2}) + 2\pi \gamma\; c(\gamma)\; 
\cot \gamma\pi\; \sin \gamma\omega  
\end{equation} 
\begin{displaymath}
\left. + \gamma \;\cos \gamma \omega \; x_{\gamma}(\omega)
 + 2\gamma^2 \sin {\gamma\omega} \; y_{\gamma}(\omega)\right\} \; ,
\end{displaymath}
where
\begin{equation}
x_{\gamma}(\omega)=\int_{0}^{\omega} d\varphi  \; \sin \gamma\varphi \;
\cot \frac{\varphi}{2}\;,\qquad y_{\gamma}(\omega)=\int_{\omega}^{\pi} 
d\varphi  \; \sin \gamma\varphi \; \ln \left( \sin {\frac{\varphi}{2}}
\right) \; .
\end{equation}

The partial wave Coulomb transition matrix $t_l(k,k^{\prime};E)$ is 
a function of three independent variables $k$, $k^{\prime}$ and $E$. 
The quantities $\xi$ and $\eta$, as well as the integration limits 
$\omega_0$ and $\omega_{\pi}$ in expression (27), also depend on those 
variables. The quantity $\kappa$ is connected with the energy $E$ by 
the formula (4). By definition (6), the Coulomb parameter $\gamma$ in 
expression (27) depends on $\kappa$ and therefore on the energy $E$. 
Note that in expression (27) for the $t$-matrix, the Coulomb interaction 
intensity $q_1 q_2$ is contained both in the factor before the integral 
and owing to the Coulomb parameter $\gamma$ [see Eq.(6)] in the terms 
in the curly braces in the integrand. \\ [.1in] 

\noindent {\bf 4. Partial wave Coulomb transition matrices for likely charged particles \\ 
at the ground bound state energy} \\ 

\noindent The expression (27) for the Coulomb transition matrix contains 
the double integration over $\varphi$ and $\omega$ and is rather complicated. 
It is easy to see that for separate values of the Coulomb parameter 
$\gamma$ (which correspond to certain energy values $E$) the integration 
over $\varphi$ and $\omega$ in Eq.(27) can be performed explicitly.
In such cases, simple analytical expressions for the pertial wave 
Coulomb $t$-matrix can be obtained. 

In particular, the integration in the expressions for  $x_{\gamma}(\omega)$ 
and $y_{\gamma}(\omega)$ (28) are simplified for integer values of the 
Coulomb parameter $\gamma = \gamma_n$ [Eq. (13)] corresponding to the 
energy spectrum of bound states of the two-particle system [Eq. (12)] with 
the energies $E=E_n$. 

Let us consider the form of the of-shell partial wave transition matrix 
for a repulsive Coulomb potential of interaction between likely charged 
particles ($q_1 q_2 >0$) at the ground bound state energy $E=E_n$. In 
this case, the Coulomb parameter, determined by the expression (13) 
with $n=1$, is equal to 
\begin{equation}
\gamma = \gamma_1 = 1\;,
\end{equation} 
the fourth term in the curly braces in Eq.(27), according to (14), is 
simplified to $\rho_1 \sin \omega$, and the integration in the fifth and 
sixth terms is carried out as follows
\begin{displaymath}
x_1(\omega)=\int_{0}^{\omega} d\varphi\sin \varphi \;\cot \frac{\varphi}{2} = 
\omega + \sin \omega\;,
\end{displaymath}
\begin{equation}
y_1(\omega)=\int_{\omega}^{\pi} d\varphi  \;\sin \varphi \; \ln \left( 
\sin {\frac{\varphi}{2}}\right) = \cos ^2 \frac{\omega}{2} - 
2 \sin ^2 \frac{\omega}{2} \;\ln \left( \sin \frac{\omega}{2} \right)\;.
\end{equation}
As a result, the formula (27) for the pertial wave Coulomb transition 
matrices (with $l=0,1,2,...)$ in the case of the repulsive interaction at 
$\gamma = 1$ (which corresponds to $E=E_1$) takes the form
\begin{displaymath}
t^r_l(k,k^{\prime};-b_1)= \frac{\pi q_1 q_2}{k k^{\prime}} \int_{\omega_{01}}^{\omega_{\pi 1}}
d\omega\; P_l \left( \frac{2\xi_1 - 1 + \cos \omega}{2\eta_1}\right) 
\end{displaymath}
\begin{equation}
\cdot \left\{ \cot \frac{\omega}{2} - \pi \cos \omega + \omega \cos \omega + 
(\rho_1 - 1) \sin \omega -2 \sin \omega\;  
\ln \left( \sin \frac{\omega}{2} \right) \right\} \;,
\end{equation} 
where, in accordance with Eq.(7),
\begin{equation}
\rho_1 = 1 - 2 \ln 2\;.
\end{equation} 

The quantities $\xi_1, \eta_1, \omega_{01}$ and $\omega_{\pi 1}$ in Eq.(31)  
are determined by the expressions for $\xi, \eta, \omega_{0}$ and $\omega_{\pi}$,  
in accordance with their definitions (23) and (25), taken at the point 
$\kappa = \kappa_1$,
\begin{displaymath}
\xi_1 = \frac{\kappa_1^2 (k^2 + {k^{\prime}}^2)}{(k^2 + \kappa_1^2)({k^{\prime}}^2 + 
\kappa_1^2)}\;,\quad \eta_1 = \frac{2 \kappa_1^2 k k^{\prime}}{(k^2 + 
\kappa_1^2)({k^{\prime}}^2 + \kappa_1^2)}\;, 
\end{displaymath}
\begin{equation}
\omega_{01} = 2 \arcsin \sqrt{\xi_1 - \eta_1}\;, \quad \; \omega_{\pi 1} = 
2 \arcsin \sqrt{\xi_1 + \eta_1}\;,
\end{equation} 

Note that the first term in the braces in the general expression (31 for 
the partial wave Coulomb transition matrix corresponds to the Born 
approximation:
\begin{equation}
t^{Born}_l(k,k^{\prime};-b_1) = v_l(k,k^{\prime}) = \frac{2\pi q_1 q_2}{k k^{\prime}} Q_l 
\left( \frac{k^2+{k^{\prime}}^2}{2 k k^{\prime}} \right)\;,
\end{equation}
where $Q_l(x)$ is the Legendre function of the second kind [18] 
\begin{equation}
Q_l(x) = \frac{1}{2} P_l(x) \ln \left( \frac{x+1}{x-1}\right)- W_{l-1}(x)\;,
\end{equation}
\begin{displaymath}
W_{-1}(x)=0\;, \qquad  W_{l-1}(x)=\sum_{k=1}^{l} \frac{1}{k} P_{l-k}(x) P_{k-1}(x)\;.  
\end{displaymath}

The fourth term in (31), which containts $\sin \omega$, because of the orthogonality 
of the Legendre polynomials
\begin{equation}
\int_{\omega_0}^{\omega_{\pi}} d\omega\; \sin \omega\; P_l \left( \frac{2\xi_1-1+ \cos \omega}
{2\eta_1} \right) = 2\eta_1 \int_{0}^{\pi} d\theta\; \sin \theta\; P_l (\cos \theta)=4\eta_1 
\delta_{l0} \;,
\end{equation}
contributes  only to the partial s-wave Coulomb $t$-matrix.

Integrating over $\omega$ in the expression (31) in the simplest case with $l=0$, 
we obtain the following formula for the partial s-wave Coulomb transition matrix 
for two likely charged particles $(q_1 q_2>0)$:
\begin{displaymath}
t^r_0(k,k^{\prime};E_1) = \frac{\pi q_1 q_2}{k k^{\prime}} \left\{ 4(\rho_1 - 1) \eta_1 - 
(2\xi_1 - 1) \ln \left(\frac{\xi_1 + \eta_1}{\xi_1 - \eta_1} \right) -2\eta_1 
\ln \left( \xi_1^2 - \eta_1^2 \right)\right. 
\end{displaymath}
\begin{equation}
\left. - \left[ \left( \pi - \omega_{\pi 1} \right) \sin \omega_{\pi 1} -  
\left( \pi - \omega_{01} \right) \sin \omega_{01} \right] \right\} \;.
\end{equation} 

Note, that in the case of the attractive Coulomb interaction ($q_1 q_2 < 0$) 
the corresponding partial s-wave Coulomb transition matrix, containing 
$\cot \gamma\pi$, has the pole singularity at the energy $E=E_1$.

Analogously, integrating over  $\omega$ in the expression (31) with $l=1$ and  
with $l=2$, we obtain the following formulas for the partial p- and d-wave 
Coulomb transition matrices in the case of the repulsive Coulomb interaction 
$(q_1 q_2>0)$:
\begin{displaymath}
t_1^r(k,k^{\prime};-b_1) = \frac{\pi q_1 q_2}{k k^{\prime}} \left\{- 1 - 
\frac{1}{\eta_1} \left[ (\xi_1^2 - \xi_1 - \eta_1^2 ) \ln \left( 
\frac{\xi_1 + \eta_1}{\xi_1 - \eta_1} \right) \right. \right.
\end{displaymath}
\begin{equation}
+\frac{1}{8} \left( \omega_{\pi 1} - \omega_{01} \right) \left( 2 \pi - \omega_{\pi 1} 
- \omega_{01} \right) 
\end{equation} 
\begin{displaymath}
+ \frac{1}{2} (2\xi_1-1)\left[ \left( \pi - \omega_{\pi 1} \right) 
\sin \omega_{\pi 1} - \left( \pi - \omega_{01} \right) \sin \omega_{01} \right] 
\end{displaymath}
\begin{displaymath}
\left. \left. + \frac{1}{8}\left[ \left( \pi - \omega_{\pi 1} \right) \sin 2 \omega_{\pi 1} - 
\left( \pi - \omega_{01} \right)\sin 2\omega_{01} \right] \right] \right\}\;,
\end{displaymath}
\begin{displaymath}
t_2^r(k,k^{\prime};-b_1) = \frac{\pi q_1 q_2}{k k^{\prime}} \left\{- \frac{1}{\eta_1} 
\left( \xi_1 + \frac{3}{2}\right) - \frac{1}{\eta_1^2} \left[ \left(\xi_1^3 - 
\frac{3}{2}\xi_1^2 - \xi_1 \eta_1^2 + \frac{1}{2}\eta_1^2 \right) 
\ln \left( \frac{\xi_1 + \eta_1}{\xi_1 - \eta_1} \right) \right. \right.
\end{displaymath}
\begin{displaymath}
+\frac{3}{16} (2\xi_1-1)\left( \omega_{\pi 1} - \omega_{01} \right) 
\left( 2 \pi - \omega_{\pi 1} - \omega_{01} \right)
\end{displaymath}
\begin{equation}
 + \left( \frac{3}{2} \xi_1^2 -  
\frac{3}{2} \xi_1 - \frac{1}{2} \eta_1^2 + \frac{21}{32}\right) 
\left[ \left( \pi - \omega_{\pi 1} \right) \sin \omega_{\pi 1} 
- \left( \pi - \omega_{01} \right) \sin \omega_{01} \right] 
\end{equation} 
\begin{displaymath}
+ \frac{3}{16} (2\xi_1-1) \left[ \left( \pi - \omega_{\pi 1} \right) 
\sin 2 \omega_{\pi 1} - \left( \pi - \omega_{01} \right)\sin 2\omega_{01} \right] 
\end{displaymath}
\begin{displaymath}
\left. \left.
+ \frac{1}{32}\left[ \left( \pi - \omega_{\pi 1} \right) \sin 3 \omega_{\pi 1} - 
\left( \pi - \omega_{01} \right)\sin 3\omega_{01} \right] \right] \right\}\;.
\end{displaymath}

Taking into account the relations
\begin{displaymath}
\cos \omega_{\pi 1} + \cos \omega_{01} = - 2(2\xi_1-1)\;,\;\;\;\;
\cos \omega_{\pi 1} - \cos \omega_{01} = - 4 \eta_1\;,
\end{displaymath}
which follow from Eq.(26), and the relations
\begin{displaymath}
\left( \pi - \omega_{\pi 1} \right) \sin 2 \omega_{\pi 1} - 
\left( \pi - \omega_{01} \right)\sin 2\omega_{01} = 2(2\xi_1 - 1)A_{-} 
-4\eta_1 A_{+}\;,
\end{displaymath}
\begin{displaymath}
\left( \pi - \omega_{\pi 1} \right) \sin 3 \omega_{\pi 1} - 
\left( \pi - \omega_{01} \right)\sin 3\omega_{01} = \left[ 4(2\xi_1 - 1)^2 
+ 16\eta_1^2 - 1 \right]A_{-} + 16(2\xi_1 - 1)\eta_1 A_{+}\;,
\end{displaymath}
where
\begin{displaymath}
A_{\pm} \equiv \left( \pi - \omega_{\pi 1} \right)\sin \omega_{\pi 1} \pm
\left( \pi - \omega_{01} \right)\sin \omega_{01}\;,
\end{displaymath}
the formulas for partial p- and d-wave component $t$-ìàòðèö³ (38) ³ (39) 
can be written in simpler forms:
\begin{displaymath}
t^r_1(k,k^{\prime};-b_1) = \frac{\pi q_1 q_2}{k k^{\prime}} \left\{- 1 - 
\frac{1}{\eta_1} \left[ (\xi_1^2 - \xi_1 - \eta_1^2 ) \ln \left( 
\frac{\xi_1 + \eta_1}{\xi_1 - \eta_1} \right) \right. \right.
\end{displaymath}
\begin{equation}
 + \frac{1}{8} \left( \omega_{\pi 1} - \omega_{01} \right) 
\left( 2 \pi - \omega_{\pi 1} - \omega_{01} \right) 
- \frac{1}{4} \left[ \left( \pi - \omega_{\pi 1} \right) 
\sin \omega_{\pi 1} \cos \omega_{01} \right.
\end{equation}
\begin{displaymath}
\left. \left. \left.
 - \left( \pi - \omega_{01} \right) 
\cos \omega_{\pi 1}\sin \omega_{01} \right] \right] \right\}\;,
\end{displaymath} 
\begin{displaymath}
t_2^r(k,k^{\prime};-b_1) = \frac{\pi q_1 q_2}{k k^{\prime}} \left\{- \frac{1}{\eta_1} 
\left( \xi_1 + \frac{3}{2}\right) - \frac{1}{\eta_1^2} \left[ \left(\xi_1^3 - 
\frac{3}{2}\xi_1^2 - \xi_1 \eta_1^2 + \frac{1}{2}\eta_1^2 \right) 
\ln \left( \frac{\xi_1 + \eta_1}{\xi_1 - \eta_1} \right) \right. \right.
\end{displaymath}
\begin{equation}
+\frac{3}{16} (2\xi_1-1)\left( \omega_{\pi 1} - \omega_{01} \right) 
\left( 2 \pi - \omega_{\pi 1} - \omega_{01} \right) 
\end{equation} 
\begin{displaymath}
+ \frac{1}{4}\left[ \left( \pi - \omega_{\pi 1} \right) \sin \omega_{\pi 1}
- \left( \pi - \omega_{01} \right) \sin \omega_{01} \right] 
\end{displaymath}
\begin{displaymath}
 \left. \left.  - \frac{1}{8} (2\xi_1-1) [ \left( \pi - \omega_{\pi 1} \right) 
\sin \omega_{\pi 1} \cos \omega_{01} 
- \left( \pi - \omega_{01} \right) 
\cos \omega_{\pi 1} \sin \omega_{01}] \right] \right\}\;.
\end{displaymath}

For comparison, we present the formulas for the corresponding partial p- and d- wave
Coulomb transition matrices, which were obtained in the case of attractive 
interaction ($q_1 q_2 < 0$) [18]:
\begin{displaymath}
t_1^a(k,k^{\prime};-b_1) = \frac{\pi q_1 q_2}{k k^{\prime}} \left\{4\xi_1 - 3 - 
\frac{1}{\eta_1} \left[ (\xi_1^2 - \xi_1 - \eta_1^2 ) \ln \left( 
\frac{\xi_1 + \eta_1}{\xi_1 - \eta_1} \right) \right. \right.
\end{displaymath}
\begin{equation}
 - \frac{1}{8} \left( \omega_{\pi 1} - \omega_{01} \right) 
\left( 2 \pi - \omega_{\pi 1} - \omega_{01} \right) 
+ \frac{1}{4} \left[ \left( \pi - \omega_{\pi 1} \right) 
\sin \omega_{\pi 1} \cos \omega_{01} \right.
\end{equation}
\begin{displaymath}
\left. \left. \left.
 - \left( \pi - \omega_{01} \right) 
\cos \omega_{\pi 1}\sin \omega_{01} \right] \right] \right\}\;,
\end{displaymath} 
\begin{displaymath}
t_2^a(k,k^{\prime};-b_1) = \frac{\pi q_1 q_2}{k k^{\prime}} \left\{ \frac{1}{\eta_1} 
\left(4 \xi_1^2-5\xi_1-\frac{8}{3} \eta_1^2 + \frac{3}{2}\right) - \frac{1}{\eta_1^2} 
\left[ \left(\xi_1^3 - 
\frac{3}{2}\xi_1^2 - \xi_1 \eta_1^2 + \frac{1}{2}\eta_1^2 \right) 
\ln \left( \frac{\xi_1 + \eta_1}{\xi_1 - \eta_1} \right) \right. \right.
\end{displaymath}
\begin{equation}
-\frac{3}{16} (2\xi_1-1)\left( \omega_{\pi 1} - \omega_{01} \right) 
\left( 2 \pi - \omega_{\pi 1} - \omega_{01} \right) 
\end{equation} 
\begin{displaymath}
- \frac{1}{4}\left[ \left( \pi - \omega_{\pi 1} \right) \sin \omega_{\pi 1}
- \left( \pi - \omega_{01} \right) \sin \omega_{01} \right] 
\end{displaymath}
\begin{displaymath}
\left. \left. + \frac{1}{8} (2\xi_1-1) [ \left( \pi - 
\omega_{\pi 1} \right) 
\sin \omega_{\pi 1} \cos \omega_{01} 
- \left( \pi - \omega_{01} \right) 
\cos \omega_{\pi 1} \sin \omega_{01} ] \right] \right\}\;.
\end{displaymath}

Taking the sign difference for $q_1 q_2$ in the coefficients before the braces 
in Eqs.(40), (41) and Eqs. (42), (43), we conclude that the expressions for the
corresponding partial wave transition matrices in the cases of attractive and 
repulsive Coulomb interaction differ only in their first terms and in the signs 
in front of the second terms. Other corresponding terms do not differ among 
themselves. \\ [.1in] 

\noindent {\bf 5. Discussion and conclusion} \\ 

\noindent The off-shell Coulomb transition matrix is directly connected 
with the Coulomb Green's function and includes all information about the 
system of interacting particles. In the previous work [18], a possibility 
to derive an analytical expression for the off-shell Coulomb transition 
matrix for two particles with the use of the Fock method of stereographic 
projection of the momentum space onto a four-dimensional unit sphere was 
studied. In the case of the attractive Coulomb between opposite charges
($q_1 q_2 < 0$), simple analytical expressions for the partial p-, d- 
and f-wave transition matrices at the ground bound state energy, i.e.
i.e. $E=E_1$, $t_l^a(k,k^{\prime};E_1)$ with $l=1, 2$ and $3$.

Note that the knowledge of the partial wave Coulomb transition matrix 
$t_{\lambda}(k,k^{\prime};E_n)$, the bound state wave function and its 
derivatives is necessary, in particular, when determining the electric 
$2^{\lambda}$-pole polarizability $\alpha_{\lambda}\quad (\lambda = 
1, 2, 3, ...)$ of a two-particle Coulomb bound system in the state with 
the energy $E=E_n$ [20]. 

In this work, the Fock method is applied in order to derive the partial 
wave Coulomb transition matrix in the case of repulsive Coulomb interaction 
(likely charged particles, $q_1 q_2)>0$) at the energy $E=E_1$. Rather 
simple analytical expressions are obtained for the partial s-, p- and d-wave 
transition matrices at the ground bound state energy, i.e. 
$t_l^r(k,k^{\prime};E_1)$ with $l=0, 1$ and $2$ [formulas (37), (40) and 
(41), respectively].

It is of interest, that in the case of particles with likely charges, for 
which bound states do not exist at all, the simplification of expressions 
for the partial wave Coulomb $t$-matrices takes place at the energies of 
the discrete spectrum of bound states for oppositely charged particles. 

It should be pointed out that the possibility to have a simple analytical 
form for the partial wave Coulomb $t$-matrix is associated with a possibility 
to carry out analytically the integration over $\varphi$ and $\omega$ in 
the expressions (28) for $x_{\gamma}(\omega)$ and $y_{\gamma}(\omega)$ and 
in the expression (27) for $t_l(k,k^{\prime};E)$. In particular, such 
integration can be done at the energy values that are equal to the energies 
of the ground and excited bound states in the discrete spectrum  $E_n$, 
$n=1,2,3, ...$ [the formula (12)]. The procedure can be realized for 
the partial wave Coulomb matrices $t_l^r(k,k^{\prime};E_n)$ that describe 
a system with repulsive forces (with likely charged particles, $q_1 q_2>0$) 
at all values of $n$ and $l$. Analytical expressions for the Coulomb 
transition matrices $t_l^a(k,k^{\prime};E)$ describing a system with 
attractive forces (with oppositely charged particles, $q_1 q_2<0$) can be 
obtained only at the values $n$ and $l$ that do not realize bound 
states at which the corresponding transition matrix has pole singularity 
(at each value of $n$ and the values of the orbital momentum with 
$l\leq n-1$). The pole singularities arise, for instance, in the partial 
wave Coulomb transition matrix $t_0^a(k,k^{\prime};E)$ at $E=E_1$, 
in the matrices $t_0^a(k,k^{\prime};E)$ and $t_1^a(k,k^{\prime};E)$  
at $E=E_2$, and so forth.

Note that the partial wave Coulomb transition matrix (27) takes the simple 
analytical form not only at the energy values corresponding to the discrete 
spectrum of bound states (12) that is equivalently, in accordance with (13), 
to integer values of the Coulomb parameter (6). A similar simplification 
can also be obtained for the Coulomb parameter value $\gamma= \frac{1}{2}$, 
which is equivalent to the negative energy $E=4E_1$. \\ [.1in]

The present work was partially supported by the National Academy of Sciences 
of Ukraine (project No. 0112U000054) and by the Program of Fundamental Research 
of the Department of Physics and Astronomy of NASU (project No. 0112000056). 
\\ [.2in]

\begin{itemize} 
\setlength{\baselineskip}{.1in} 
\item[{\tt 1.}] J.C.Y.Chen and A.C.Chen, in {\it Advances in Atomic and Molecular 
                Physics}{\bf 8}, edited by D.B.Bates and I.Estermann (Academic Press, 
                New York, London, 1972) p.p. 71-129.
\item[{\tt 2.}] L.D.Faddeev, Sov. Phys. JETP {\bf 12}(1961)1014-1019.
\item[{\tt 3.}] L.D.Faddeev, Mathematical Aspects of the Three-Body Problem in the 
                Quantum Scattering Theory. Isr. Program Sci. Transl., Jerusalem, 1965. 
\item[{\tt 4.}] O.A.Yakubovsky Sov. J. Nucl. Phys. {\bf 5}(1967)937-942.
\item[{\tt 5.}] A.M.Veselova, Teor. Mat. Fiz. {\bf 3}(1970)326-331.
\item[{\tt 6.}] V.G.Gorshkov, Zh. Eksp. Teor. Fiz. {\bf 40}(1961)1481-1490.
\item[{\tt 7.}] G.Ya.Beil'kin, Vest. Leningrad. Univ. No.{\bf 13}(1978)72.
\item[{\tt 8.}] S.Okubo and D.Feldman, Phys. Rev. {\bf 117}(1960)292-306.
\item[{\tt 9.}] E.H.Wichmann and C.H.Woo, J. Math. Phys. {\bf 2}(1961)178-181.
\item[{\tt 10.}] V.F.Bratsev and E.D.Trifonov, Vest. Leningrad. Univ. No.{\bf 16}(1962)36-39. 
\item[{\tt 11.}] L.Hostler, J. Math. Phys. {\bf 5}(1964)1235-1240.
\item[{\tt 12.}] J.Schwinger, J. Math. Phys. {\bf 5}(1964)1606-1608.
\item[{\tt 13.}] A.M.Perelomov and V.S.Popov, Sov. Phys. JETP {\bf 23}(1966)118-134.
\item[{\tt 14.}] S.A.Shadchin and V.F.Kharchenko, J. Phys. B: At. Mol. Phys. {\bf 16}(1983)1319-1322.  
\item[{\tt 15.}] S.A.Storozhenko and S.A.Shadchin, Teor. Mat. Fiz. {\bf 76}(1988)339-349.  
\item[{\tt 16.}] H.vanHaeringen, J. Math. Phys. {\bf 25}(1984)3001-3032. 
\item[{\tt 17.}] V.A.Fock, Z. Phys. {\bf 98}(1935)145-154.  
\item[{\tt 18.}] V.F.Kharchenko, Annals of Physics {\bf 374}(2016)16-26.  
\item[{\tt 19.}] I.S.Gradstein, I.M.Ryzhik, Tables of Integrals, Series, and Products
                (Academic,1980).  
\item[{\tt 20.}] V.F.Kharchenko, J. Mod. Phys. {\bf 4}(2013)99-107. 

\end{itemize} 

\end{document}